\documentclass[conference]{IEEEtran}
\IEEEoverridecommandlockouts
\pdfoutput=1
\usepackage{cite}
\usepackage{amsmath,amssymb,amsfonts}
\usepackage{algorithmic}
\usepackage{graphicx}
\usepackage{epstopdf}
\usepackage{graphicx}
\usepackage{textcomp}
\usepackage{xcolor}
\def\BibTeX{{\rm B\kern-.05em{\sc i\kern-.025em b}\kern-.08em
    T\kern-.1667em\lower.7ex\hbox{E}\kern-.125emX}}
\begin{document}
\pdfoutput=1
\title{Sparse and Safe Frequency Regulation for Inverter Intensive Microgrids\\
\thanks{This material is based upon work supported by the U.S. Department of Energy's Office of Energy Efficiency and Renewable Energy (EERE) under the Solar Energy Technologies Office Award Number 38637. The views expressed herein do not necessarily represent the views of the U.S. Department of Energy or the United States Government.}
}

\author{\IEEEauthorblockN{Junhui Zhang, Lizhi Ding, Xiaonan Lu}
\IEEEauthorblockA{\textit{College of Engineering} \\
\textit{Temple University}\\
Philadelphia, USA \\
\{junhui.zhang, lizhi.ding, xiaonan.lu\}@temple.edu}
\and
\IEEEauthorblockN{Wenyuan Tang}
\IEEEauthorblockA{\textit{Department of Electrical and Computer Engineering} \\
	\textit{North Carolina State University}\\
	Raleigh, USA  \\
	wtang8@ncsu.edu}
}

\maketitle

\begin{abstract}
This paper developed a novel control approach for the sparse and safe frequency regulation for inverter intensive microgrids (MGs). In the scenario, the inverters and external grids are expected to reach a synchronized desired frequency under regulations. To this end, the active power set-point acting as a control from a high-level controller is designed while considering two important performance metrics ``sparsity" and ``safety", which are to reduce the information exchange between controllers and ensure that the frequency keeps in safety regions during the whole operation process. Our proposed control design framework allows the sparse linear feedback controller (SLFC) to be unified with a family of conditions for safe control using control barrier functions. A quadratic programming (QP) problem is then constructed, and the real-time control policy is obtained by solving the QP problem. Importantly, we also found that the real-time control for each inverter depends on the cross-layer communication network topology which is the union of the one between controllers from SLFC and the one determined by the power flow network. Furthermore, our approach has been validated through extensive numerical simulations.  
\end{abstract}

\section{Introduction} 
Microgrids (MGs) were proposed and developed  with the rapid penetration
of renewable energy resources \cite{guerrero2010hierarchical}. As a promising way to cope with the intermittency and uncertainty of renewables, the research on MGs draws increasing attention and plays an essential role in the area of smart grids. In particular, frequency regulation, i.e., a desired frequency is expected under regulations, is one of the most challenging and vital control requirements in MGs. Up to now, there have been tremendous efforts \cite{simpson2015secondary,qian2019event,alghamdi2020synthesizing,kundu2019distributed}, either in centralized or distributed  manner. However, as far as we know, the frequency regulation in MGs considering the two important performance metrics ``sparsity" and ``safety" simultaneously has been largely ignored in the existing literature.

The consideration of ``sparsity" is based on the fact that information exchange between controllers is necessary to achieve the synchronized desired frequency, i.e., the system-level regulation. While communication networks support the information exchange among controllers in MGs, the threat from malicious cyber attacks is also introduced to the MGs and thus serious damage may be triggered \cite{liu2018stochastic,lu2021communication,zhang2021optimal}. On the other hand, the communication network bears huge burden when information exchange is processed heavily. To enhance the resilience of MGs and reduce the burden of information transmission, the performance metrics  ``sparsity" should be taken into consideration when a control policy is designed. This is also named as sparse control, in which the number of communication links between controllers is reduced while guaranteeing other control performance. The algorithm for sparse control was developed in \cite{lin2013design} via the alternating direction methods of multipliers (ADMM). The applications of sparse control were seen from the wide-area power networks \cite{dorfler2014sparsity,dizche2019sparse},  voltage and current regulations for DC MGs \cite{liu2019secondary} to inter-area oscillations \cite{wu2015input}.  

 Due to the low-inertia nature of inverter intensive MGs, large frequency fluctuations are commonly observed. Thus, when we design a control policy for frequency regulation, it is important to ensure that the frequency always satisfies the ``safety" constraints \cite{smith2021transient}. That is, the frequency should not deviate from its desired value too much during the whole process.  Recently, the control barrier functions (CBFs) have become a promising tool to design safe control for safety-critical systems, such as collision avoidance of connected automated vehicles \cite{xiao2019decentralized} and trajectory planning of robotics \cite{ames2016control}. This motivates us to develop a CBF-based safe controller for frequency regulation in MGs. Moreover, how to enhance sparsity for the safe controller while guaranteeing stability has also not been addressed in literature, which is the technical gap to be filled in this work.    
 
In this paper, we develop a novel control approach for the sparse and safe frequency regulation for inverter intensive MGs. The framework of this control design is comprised of three steps. First, the sparse linear feedback controller (SLFC) is obtained by solving a sparsity-promoting optimal control problem, which acts as a nominal control. Second, the CBFs are applied to design a family of conditions for safe control that guarantees the satisfaction of frequency safety constraints. Finally, by unifying SLFC and CBFs, a quadratic programming (QP) problem is constructed, which minimizes the difference between the real control and the nominal control while ensuring the conditions resulting from CBFs hold. The real-time control is then obtained by solving the QP problem.

Compared with the existing work, our contributions in this work are summarized:
\begin{itemize}
	\item A novel control approach for the sparse and safe frequency regulation  in inverter intensive MGs is developed. 
	%Different from the works \cite{dorfler2014sparsity,liu2019secondary,lin2013design,kundu2019distributed,bouvier2022distributed,smith2021transient}, in which only ``sparse" or only ``safe" control is designed, 
	The proposed control approach unifies the two performance metrics ``sparsity" and ``safety" by constructing a QP problem and a real-time control policy is obtained by solving the QP problem. 
	\item We found that the real-time control for each inverter depends on the cross-layer communication network topology which is the union of the one between controllers from SLFC and the one determined by the power flow network.  
    \item The framework of our sparse and safe control approach has remarkable applicability and holds great potentials to address other problems in power systems. 
\end{itemize}

The remainder of this paper is organized as follows.  In Section II, some preliminaries are given. Section III presents the model of MGs with inverters 
and states the problem of interest. In Section IV, the sparse and safe control approach is proposed. In Section V, the case study is given to verify our control approach. Finally, some conclusions are given in Section VI.

\section{Preliminaries} 
In this section, we give some preliminaries for the control barrier function (CBF), which is a powerful tool to design a safe controller. Consider the affine nonlinear system:
\begin{align}
\dot{x}=f(x)+g(x)u,
\end{align}
where functions $f$ and $g$ are locally Lipschitz continuous. 
$x\in R^n$ and $u\in R^m$ are state and control of the system (1), respectively.

\textbf{Definition 2.1}\cite{ames2016control}. A continuous function $\alpha: [-b, a)\rightarrow [-\infty, +\infty)$ is said to be an extended class $K$ function for $a,b>0$ if it is strictly increasing and $\alpha(0)=0$.

\textbf{Definition 2.2} \cite{xiao2019decentralized}. A set $\Omega$ is forward invariant for system (1) if its solutions starting at all $x(t_0)\in \Omega$ satisfy $x(t)\in \Omega$ for $\forall t\geq t_0$.

\textbf{Definition 2.3}\cite{ames2016control}. Given a set $\Omega=\{x\in R^n: h(x)\geq 0\}$ for a continuously differentiable function $h:R^n\rightarrow R$, the function $h$ is called a control barrier function, if there exists an extended class $K$ function $\alpha$ such that
\begin{align*}
\sup \limits_{u\in R^m}\{L_fh(x)+L_gh(x)u+\alpha(h(x))\}\geq 0,
\end{align*}
where $L_fh(x)=\frac{\partial h}{\partial x}f(x)$ and $L_gh(x)=\frac{\partial h}{\partial x}g(x)$ are Lie derivatives, $\alpha(\cdot)$ is an extended class $K$ function.

 For a CBF $h(x)$, define the set for $x\in R^n$,
 \begin{align*}
 K_{cbf}(x)=\{u\in R^m: L_fh(x)+L_gh(x)u+\alpha(h(x))\geq 0\}.  \end{align*}
 Then, any Lipschitz continuous controller $u\in K_{cbf}(x)$, will render the set $\Omega=\{x\in R^m: h(x)\geq 0\}$ forward invariant for control system (1)\cite{ames2016control}. 

  \section{Microgrid Models and  Problem Formulation} 
\subsection{Inverter intensive microgrids}
\begin{figure}[!htb]
	\begin{center}
		\includegraphics[width=2.8in]{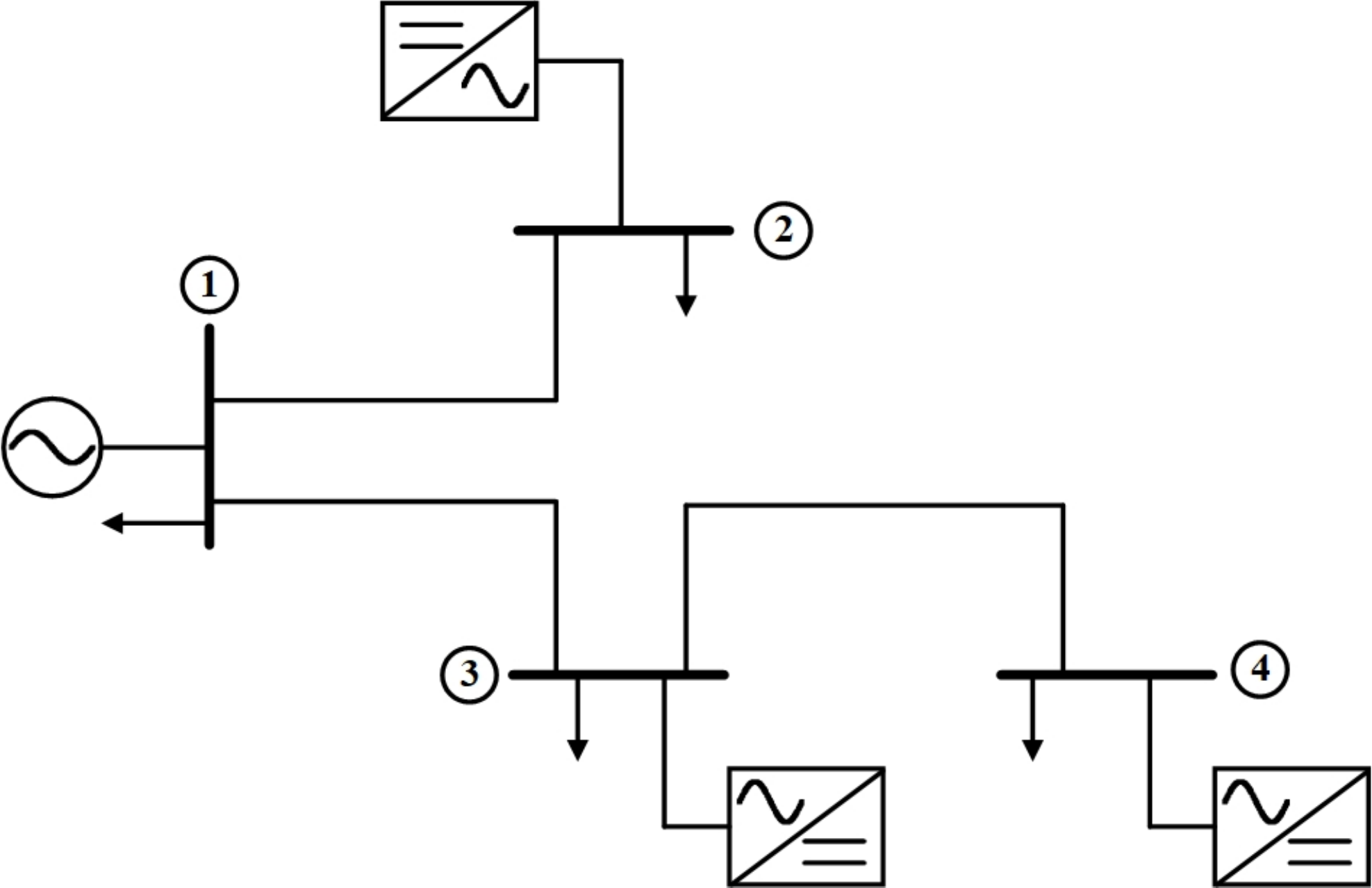}
		\hspace{100in}\vspace{-3mm}
		\caption{Inverter intensive MGs.}
	\end{center}
\end{figure}
In this work, we focus on the MGs with inverters. From \cite{schiffer2014conditions}, the dynamic of droop-controlled inverter is well constructed, which shares the same mathematical model with synchronous generator (SG). We describe the inverters in MGs by nodes of a graph. Then, for each node $i$, one has  
\begin{align}
\dot{\theta}_i&= \omega_i, \\
M_i\dot{\omega}_i&=-D_i(\omega_i-\omega^d)+P_i^{set}-P_i,
\end{align}
where $\theta_i$, $\omega_i$ and  $w^d$ are the phase angle, 
frequency and the desired frequency, respectively. $M_i=\tau_i/\lambda_i^p$, $D_i=1/\lambda^p_i$, where $\lambda_i^p$ represents the droop coefficient and $\tau_i>0$ is the time constant of the power measurement filter (PMF). In this work, we are concerned with frequency regulation for the MGs, thus it is rational to assume 
the voltage amplitudes at all nodes are positive real constants \cite{schiffer2019global}. Additionally, suppose that the considered MGs are with dominantly inductive admittances. $P_i^{set}$ is the active power set-point. Then the active power flow $P_i$ injected into the network is given by \cite{schiffer2019global}
\begin{align}
P_i&=G_{i,i}v_i^2-\sum_{j\in\mathcal{N}_i}v_iv_j\mathcal{B}_{i,j}sin\theta_{i,j},
\end{align}
where $\theta_{i,j}=\theta_i-\theta_j$. $v_i$ and $v_j$ are the voltage amplitudes at node $i$ and $j$, respectively. $\mathcal{N}_i$ is the set of neighbors of node $i$ which is described as $\mathcal{N}_i=\{j| j\neq i, B_{i,j}\neq 0\}$ .  $B_{i,j}$ is the transfer susceptance value of the line between node $i$ and $j$. $G_{ii}$ is the shunt conductance at node $i$, i.e., the active power load.

In order to shift the desired frequency to zero, we define $\hat{\omega}_i=\omega_i-\omega^d$. Thus the systems (2) and (3) are rewritten as 
\begin{align}
\dot{\theta}_i&=\hat{\omega}_i+\omega^d, \\
M_i\dot{\hat{\omega}}_i&=-D_i\hat{\omega}_i+P_i^{set}-P_i,
\end{align}
with the active power $P_i$ in (4). Furthermore, let $\hat{\theta}_i=\theta_i-\theta_{i}^d=\theta_i-\theta_{i,0}-\omega^dt$, where $\theta_{i,0}$ is a constant. Then
\begin{align}
\dot{\hat{\theta}}_i=&\hat{\omega}_i,\\
\nonumber
M_i\dot{\hat{\omega}}_i=&-D_i\hat{\omega}_i+P_i^{set}-G_{i,i}v_i^2\\&+\sum_{j\in\mathcal{N}_i}v_iv_j(\mathcal{B}_{i,j}sin(\hat{\theta}_{i,j}+\theta_{i,0}-\theta_{j,0})).
\end{align}
From \cite{schiffer2014conditions}, the active power set-point $P_i^{set}$ is from a high-level controller and acts as a control signal to the systems (7)-(8).
\subsection{Approximation}

In practice, the angle differences between nodes are usually small. Thus, it is rational to apply the approximation $sin(\hat{\theta}_{i,j}+\theta_{i,0}-\theta_{j,0})\approx\hat{\theta}_{i,j}+\theta_{i,0}-\theta_{j,0}$. The dynamics (7)-(8) are rewritten as
\begin{align}
\nonumber
\begin{bmatrix} \dot{\hat{\theta}}_i\\ \dot{\hat{\omega}}_i \end{bmatrix}=&\begin{bmatrix} 0 & 1\\ 0 & -D_i/M_i \end{bmatrix}\begin{bmatrix} \hat{\theta}_i\\ \hat{\omega}_i \end{bmatrix}+\begin{bmatrix}0\\M_i^{-1}\end{bmatrix}P^{set}_i\\
&+
\begin{bmatrix} 0 & \cdots & 0\\ \frac{v_iv_1\mathcal{B}_{i,1}}{M_i}&\cdots&\frac{v_iv_n\mathcal{B}_{i,n}}{M_i}\end{bmatrix}\begin{bmatrix} \hat{\theta}_{i,1}\\ \vdots\\ \hat{\theta}_{i,n}\end{bmatrix}+\begin{bmatrix} 0 \\ M_i^{-1} \end{bmatrix}c_i,
\end{align}
where $c_i$ is a constant. Thus, the final term in (9) can be compensated easily by designing $P_i^{set}=u_i-c_i$. Also, we further consider the fact that the loads in MGs may change with time. This results in an un-negligible exogenous disturbance to the dynamics. 
Let $\hat{\theta}=col(\hat{\theta}_i)$, $\hat{\omega}=col(\hat{\omega}_i)$, $x=[\hat{\theta}^T \ \ \hat{\omega}^T]^T,$ the linear state-space equation is given by
\begin{align}
\dot{x}&=Ax+B_2u+B_1d,
\end{align}
where $A=\begin{bmatrix}0&I\\-M^{-1}L&-M^{-1}D\end{bmatrix}$, $B_2=\begin{bmatrix}0\\M^{-1}\end{bmatrix}$, $B_1=\begin{bmatrix}0\\M^{-1}V^2\end{bmatrix}$, $M=diag\{M_1,\cdots, M_n\}$, $D=diag\{D_1,\cdots, D_n\}$, $V=diag\{v_1,\cdots, v_n\}$.
$u$ is the control and $d$ is the exogenous disturbance resulting from load fluctuations. $L$ denotes the Laplacian matrix, and its entries are given by
\begin{align*}
L_{ii}&=-\sum_{j=1,j\neq i}^{n}\mathcal{B}_{i,j}v_iv_j, \ \ \text{if} \ \ i=j,\\
L_{i,j}&=\mathcal{B}_{i,j}v_iv_j, \ \ \text{if}\ \ i\neq j.
\end{align*}
\subsection{Problem formulation}
In this work, we are to design a control approach for frequency regulations in inverter intensive MGs while considering sparsity and safety. For the sparsity, the information exchange should be reduced effectively, while stabilizing the system (10). For the safety, 
the frequency should not deviate from its desired value too much during the whole process. Specifically, the safety constraints for frequency are given by
\begin{align}
-\omega_{l}\leq\hat{\omega}_i\leq \omega_{h}
\end{align}
with $\omega_{l}>0$ and $\omega_{h}>0$, which define the safety regions for frequency. In view of the discussion above, the goal of this work is to design a sparse and safety control $u$
such that 
\begin{itemize}
	\item The information exchange between controllers are reduced while stabilizing the system (10).
    \item The frequency safety constraints (11) are always satisfied during the whole operation process.   
\end{itemize}

\section{Our proposed sparse and safe control approach} 

\subsection{Sparsity-promoting optimal control}

The closed-loop system (10) with linear feedback control  is written as 
\begin{align}
\dot{x}&=Ax+B_2u+B_1d,\\
u&=-Kx,\\
z&=\begin{bmatrix}Q^{1/2}\\-R^{1/2}K\end{bmatrix}x,
\end{align}
where $K$ is the linear feedback gain and $z$ is the performance output.  $Q=Q^T\geq0$ and $R=R^T>0$ are the weighting matrices for the state and control performance, respectively. $(A, B_2)$ is stabilizable and $(A,Q^{1/2})$ is detectable.

Typically, the $H_2$ optimal control for system (12) is to find the optimal feedback gain $K^*$, which minimizes the $H_2$ norm of the transfer function from $d$ to $z$ while stabilizing the system (12). The standard $H_2$ optimal control problem is described as
\begin{align}
\nonumber 
\text{minimize} &\ J(K)=Tr(B_1^TP(K)B_1),
\end{align}
where $Tr(\cdot)$ is the trace of a matrix, the closed-loop observability Gramian $P(K)$ is 
\begin{align}
P(K)=\int_0^{\infty}e^{{(A-B_2K)}^Tt}(Q+K^TRK)e^{(A-B_2K)t}dt,
\end{align}
which is the solution of $(A-B_2K)^TP+P(A-B_2K)=-(Q+K^TRK).$

In order to enhance the sparsity of controller, we add a penalty in the objective function. Thus, the sparsity-promoting optimal control problem is constructed as
\begin{align}
\text{minimize} &\ J_{\gamma}(K)=Tr(B_1^TP(K)B_1)+\gamma Card(K),
\end{align}
where $Card(\cdot)$ represents the cardinality function and $\gamma\geq 0$ is the weighting coefficient.
Typically, the weighted $l_1$ norm is used to replace the cardinality function in the optimization problem \cite{boyd2004convex,lin2013design},
\begin{align}
Card(K)=\sum_{i,j}W_{i,j}|K_{i,j}|,
\end{align}
and the weights $W_{i,j}$ are chosen as
\begin{eqnarray}
W_{i,j}=
\frac{1}{|K_{i,j}|+\epsilon},
\end{eqnarray}
where $\epsilon>0$ is a small constant. Hence, the sparsity-promoting optimal control problem (16) is rewritten as 
\begin{align}
\text{minimize} &\ J_{\gamma}(K)=Tr(B_1^TP(K)B_1)+\gamma \sum_{i,j}W_{i,j}|K_{i,j}|.
\end{align}
To solve the sparsity-promoting optimal control problem (19), we can use the alternating direction method of multipliers (ADMM) developed in \cite{lin2013design}.

\subsection{Safety Guarantees-Control Barrier Function} 

During the operation process, the frequency of each inverter should keep in safe regions, which are described by the constraints (11). Define functions $h_{1,i}\triangleq \hat{\omega}_i+\omega_l$, $h_{2,i}\triangleq \omega_h-\hat{\omega}_i$. Hence, to guarantee the safety of frequency is equivalent to ensuring that the following inequalities hold,
\begin{align}
h_{1,i}\geq0,\ \ h_{2,i}\geq0,
\end{align}
which result in the set $\Omega_{i}\triangleq\{(\hat{\theta}_i,\hat{\omega}_i )|h_{1,i}\geq0,\ \ h_{2,i}\geq0\}$. Thus when $\Omega_i$ is forward invariant, the frequency safety constraints are satisfied. Consider the control barrier functions $h_{1,i}$, $h_{2,i}$, the derivative of $h_{1,i}$ is given by
\begin{align}
\dot{h}_{1,i}= -M_i^{-1}D_i\hat{\omega}_i+\sum_{j\in\mathcal{N}_i} \frac{v_iv_j\mathcal{B}_{i,j}}{M_i}\hat{\theta}_{i,j}+M_i^{-1}u_i+M_i^{-1}v_i^2d_i,
\end{align}
in which we see that $h_{1,i}$ has relative degree one. Note $\dot{h}_{2,i}=-\dot{h}_{1,i}$, thus $h_{2,i}$ also has relative degree one. To ensure the set $\Omega_i$ is forward invariant, the extended class $K$ functions $\alpha_1$ and $\alpha_2$ are selected as linear functions, that is, $\alpha_1: h_{1,i}\rightarrow \eta_1 h_{1,i}$, $\alpha_2: h_{2,i}\rightarrow \eta_2h_{2,i}$, where $\eta_1>0$, $\eta_2>0$. Hence, the condition for safe control is given by
\begin{align}
\dot{h}_{1,i}+\eta_1h_{1,i}\geq0, \ \ \dot{h}_{2,i}+\eta_2h_{2,i}\geq0.
\end{align}
Note that the condition (22) contains the disturbance $d_i$, which is unknown for us. So the condition (22) can not be used to search for the safe control directly. Also, we note that the load fluctuation in the real world is not unlimited unless it meets extreme disasters. Thus, it is reasonable to assume that the disturbance is bounded by a positive constant $d_s$, that is, $|d_i|\leq d_s$. Then the conditions for safe control resulting from robust control barrier functions are given by
\begin{align}
\nonumber
&-M_i^{-1}D_i\hat{\omega}_i+\sum_{j\in\mathcal{N}_i} \frac{v_iv_j\mathcal{B}_{i,j}}{M_i}\hat{\theta}_{i,j}+M_i^{-1}u_i-M_i^{-1}v_i^2d_s\\
&+\eta_1\hat{\omega}_i+\eta_1\omega_l\geq 0,\\
\nonumber
&M_i^{-1}D_i\hat{\omega}_i-\sum_{j\in\mathcal{N}_i} \frac{v_iv_j\mathcal{B}_{i,j}}{M_i}\hat{\theta}_{i,j}-M_i^{-1}u_i-M_i^{-1}v_i^2d_s\\
&+\eta_2\omega_h-\eta_2\hat{\omega}_i\geq 0.
\end{align}
\subsection{QP-based control}

After using the ADMM, a sparse linear feedback gain $K^*$ is obtained, then $-K^*x$ acts as the nominal control. By enforcing the conditions (23) and (24),  a QP problem by minimizing the difference between the real control and the nominal control is constructed, 
\begin{align}
\nonumber
u_i^*=&\mathop{\arg}\min\limits_{u_i} \ \|u_i-u_i^0\|_2\\
s.t. \ \  &\textbf{A}_{cbf}u_i\leq \textbf{b}_{cbf}
\end{align}
where $u_i^0=-Row(K^*)_ix$,
$\textbf{A}_{cbf}$=$
\begin{bmatrix}
-M_i^{-1}\\
M_i^{-1}
\end{bmatrix}$,\\
$\textbf{b}_{cbf}$=\\$
\begin{bmatrix}
-M_i^{-1}D_i\hat{\omega}_i+\sum\limits_{j\in\mathcal{N}_i} \frac{v_iv_j\mathcal{B}_{i,j}}{M_i}\hat{\theta}_{i,j}-M_i^{-1}v_i^2d_s+\eta_1\hat{\omega}_i+\eta_1\omega_s\\
M_i^{-1}D_i\hat{\omega}_i-\sum\limits_{j\in\mathcal{N}_i} \frac{v_iv_j\mathcal{B}_{i,j}}{M_i}\hat{\theta}_{i,j}-M_i^{-1}v_i^2d_s+\eta_2\omega_s-\eta_2\hat{\omega}_i
\end{bmatrix}$.\\
\begin{figure}[!htb]
	\begin{center}
		\includegraphics[width=3in]{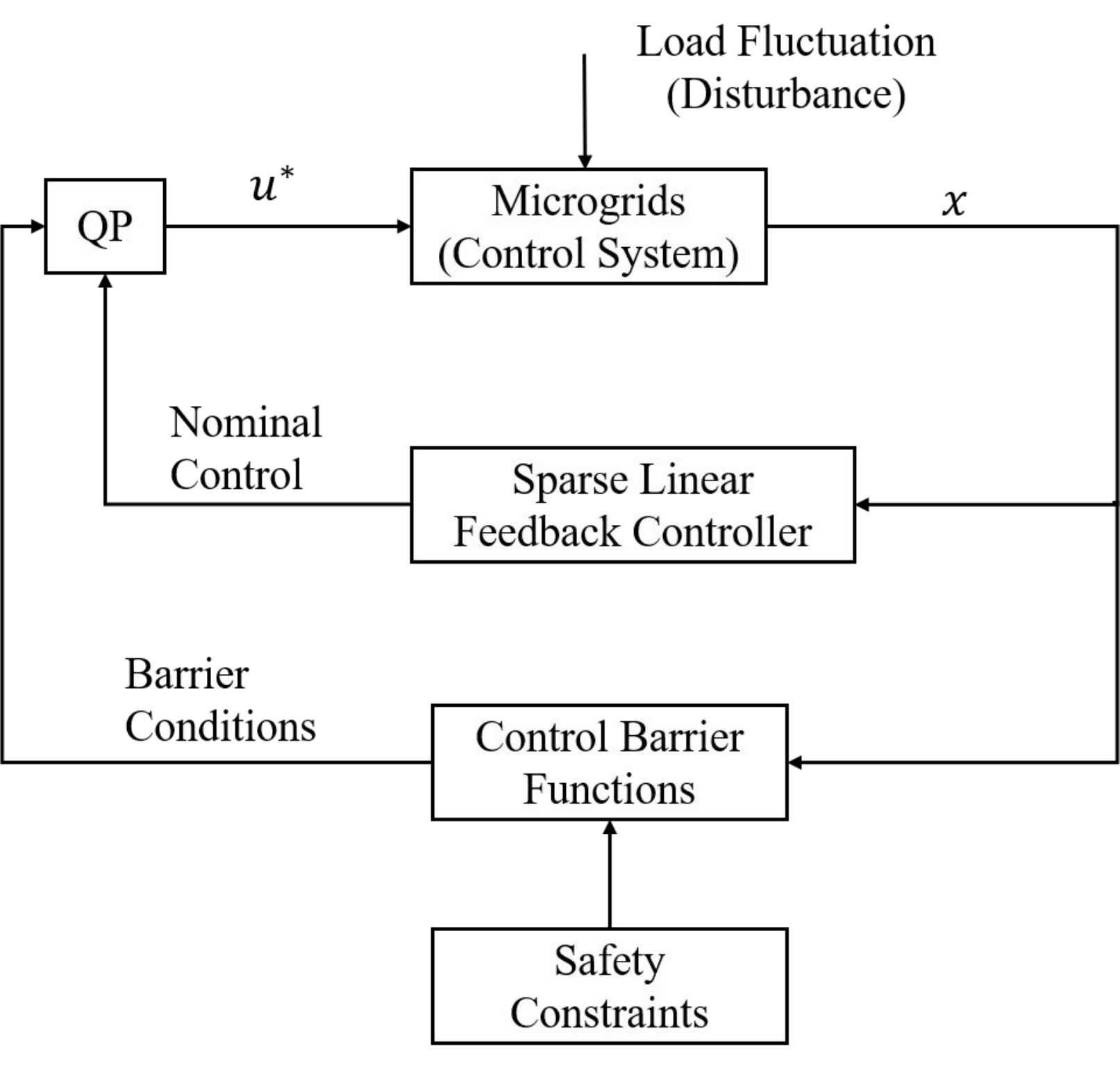}
		\hspace{100in}\vspace{-3mm}
		\caption{The design framework of the sparse and safe control.}
	\end{center}
\end{figure}

Then, the real-time control is obtained by solving the QP (25).
In addition, the design framework of the sparse and safe control is depicted as the flow chart in Fig. 2.

Especially, in the QP (25) for each inverter, the nominal control $u_i^0$ needs the information exchange by the SLFC and the $\textbf{b}_{cbf}$ needs the information from its neighbors determined by the power flow network. This implies that the real-time control for each inverter depends on the cross-layer communication network topology which is the union of the one between controllers from SLFC and the one determined by the power flow network.  

\section{Case Study}
In this section, a case study is given to verify our control approach. 

\begin{table}[htbp]
	\centering
	\caption{Electrical and Droop Control Parameters}
	\begin{tabular}{lllll}     \hline
		\textbf{Parameters}& \textbf{Value}\\ \hline
	    Nominal frequency&  60 Hz \\ 
		Inertia constant of SG (Bus 1)&  0.4kg $\cdot$ $\text{m}^2$\\  
		Damping coefficient of SG (Bus 1)& 0.8\\
		Droop coefficient of inverter (Bus 2) & 2.43 Hz/p.u.\\  
		Droop coefficient of inverter (Bus 3) & 2.43 Hz/p.u.\\
		Droop coefficient of inverter (Bus 4) & 2.43.Hz/p.u. \\
		Filter time-constant of inverter (Bus 2) & 0.5s \\   
		Filter time-constant of inverter (Bus 3) & 0.5s \\
		Filter time-constant of inverter (Bus 4) & 0.5s \\
		Line impedance (1,2)&  $(2.5+j8.7)10^{-3}$ \\  
        Line impedance (1,3)&  $(2.5+j8.7)10^{-3}$ \\  
		Line impedance (3,4)&  $(2.5+j8.7)10^{-3}$ \\ \hline
	\end{tabular}
\end{table}

We consider a 4-bus (bus 1-bus 4) MG, in which bus 1 is a SG and the others are droop-controlled inverters. The power flow network is shown in Fig. 1. The dynamics of inverters and SG can be modeled by eqs. (2)-(3). For the SG on bus 1, the values of inertia constant and  damping coefficient are shown in Table 1. For inverters on bus 2-bus 4, the values of droop coefficients $\lambda_i^p$ and filter time-constants $\tau_i$ are also given in Table 1. From \cite{ersal2011impact}, the line impedance between two connected buses is $(2.5+j8.7)10^{-3}$. The voltage for each bus is 1 p.u.. The synchronized desired frequency is 60 Hz (The standard in the United States). By shifting the desired frequency to zero and approximating the sine function in the power flow, the linear state equation is given as (10). To penalize the deviation of the angle and frequency, we take $Q=I_8$. 
Also, to quantify the control cost, the weighting matrix $R$ is taken as $I_4$. Then, the sparsity-promoting optimal control problem (18) is solved for 50 logarithmically spaced values of $\gamma\in[10^{-4}, 10^{-1}]$. The simulation is done with the software in $www.umn.edu/~mihailo/software/lqrsp/$ developed in \cite{lin2013design}, and the results are depicted as Fig. 3 and Fig. 4. In theory, when $\gamma=0$, the sparsity-promoting optimal control is degenerated into the standard $H_2$ control and the control policy is in the centralized manner. As we can see in the Fig. 3, when $\gamma$ is very small, i.e., $\gamma\leq 0.0026$, $Card(K^*)=32$ which is corresponding to the centralized control policy. As $\gamma$ increases, the feedback matrix $K^{*}$ becomes more and more sparser. When $\gamma=10^{-1}$, the $Card(K^*)$ has been decreased to 11. Fig. 4 shows the structure of the feedback gain $K^*$ when $Card(K^*)=32$, $26$, and $11$, respectively. The blue point indicates that the element at corresponding position of $K^*$  is nonzero, and the empty means the one at this position is zero.

\begin{figure}[!htb]
	\begin{center}
		\includegraphics[width=3in]{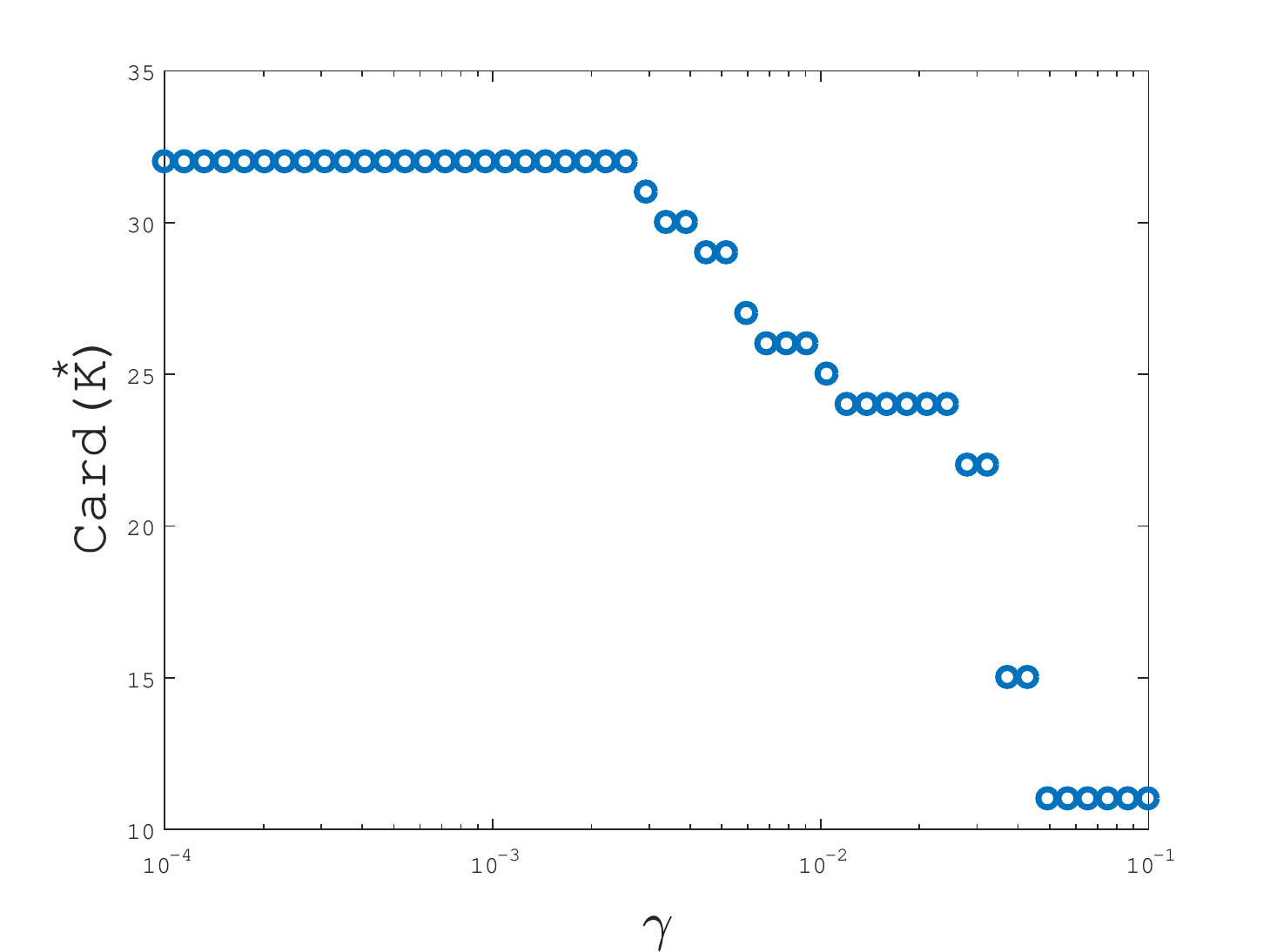}
		\hspace{80in}\vspace{-3mm}
		\caption{$Card(K^*)$.}
	\end{center}
\end{figure}
\begin{figure*}[!htb]
	\begin{center}
		\includegraphics[width=7in]{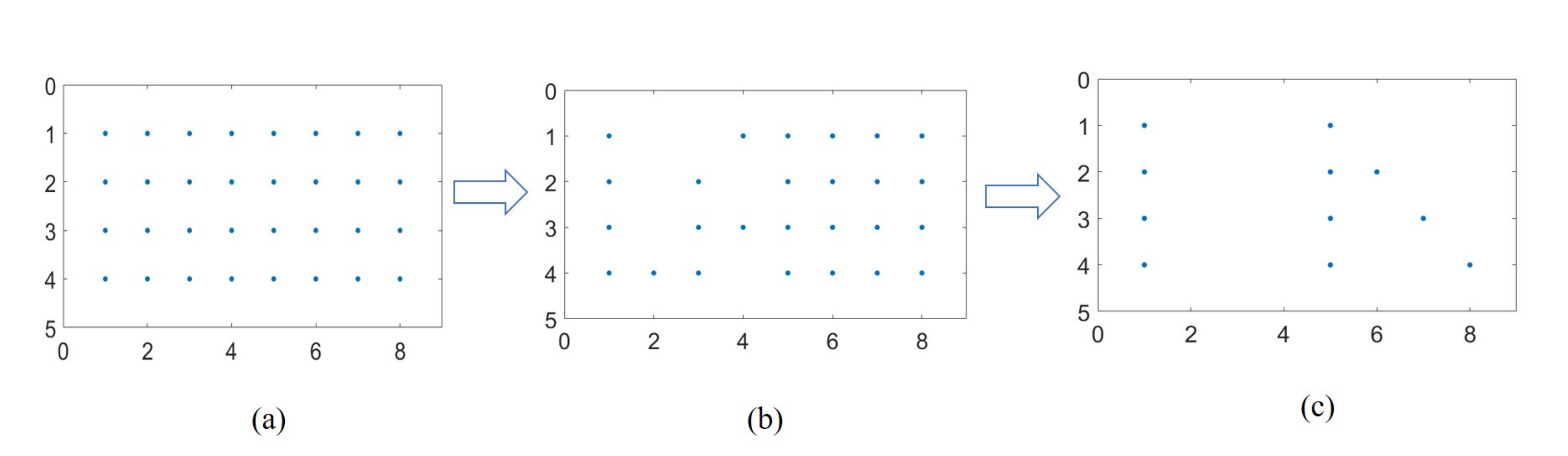}
		\hspace{100in}\vspace{-3mm}
		\caption{The structure of the feedback gain $K^*$ (a) $Card(K^*)=32$, (b) $Card(K^*)=26$, (c) $Card(K^*)=11$.}
	\end{center}
\end{figure*}

By considering the safe constraints for frequency, $\omega_{l}$ and $\omega_{h}$ are taken as $0.5$ Hz. That is,  $-0.5 \text{Hz}\leq\hat{\omega}_i\leq 0.5 \text{Hz}$ should be always satisfied during the whole operation process. Then the CBFs are constructed as (20). We assume that the disturbance is bounded by $0.5$. The conditions for safety control resulting from robust CBFs are given by (23) and (24). We take the sparse linear feedback gain $K^*$ with $Card(K^*)=11$ and regard the sparse feedback control $-K^*x$ as the nominal control. By enforcing the conditions (23) and (24), the QP problem is constructed as (25). Then the real-time control is obtained by solving the QP.
We take the initial values of $\hat{\theta}_i$ from $[0,\pi/2]$ randomly, and take initial values of $\hat{\omega}_i$ from $[-0.5,0.5]$ randomly. The sampling period $\Delta t$ is chosen as $10^{-3}s$. At each sampling time $t_k=k\Delta t$, the control is obtained by solving QP (25). The simulation is done with matlab and the results are depicted as Fig. 5 and Fig. 6. From Fig. 5, we see that $\hat{\theta}_i$ converge to zero. Importantly, from Fig. 6, we see that all $\hat{\omega}_i$ are stabilized to zero and they always keep in the safety regions during the whole process. The results illustrate that our proposed sparse and safe control approach works well.
\begin{figure}[!htb]
	\begin{center}
		\includegraphics[width=3in]{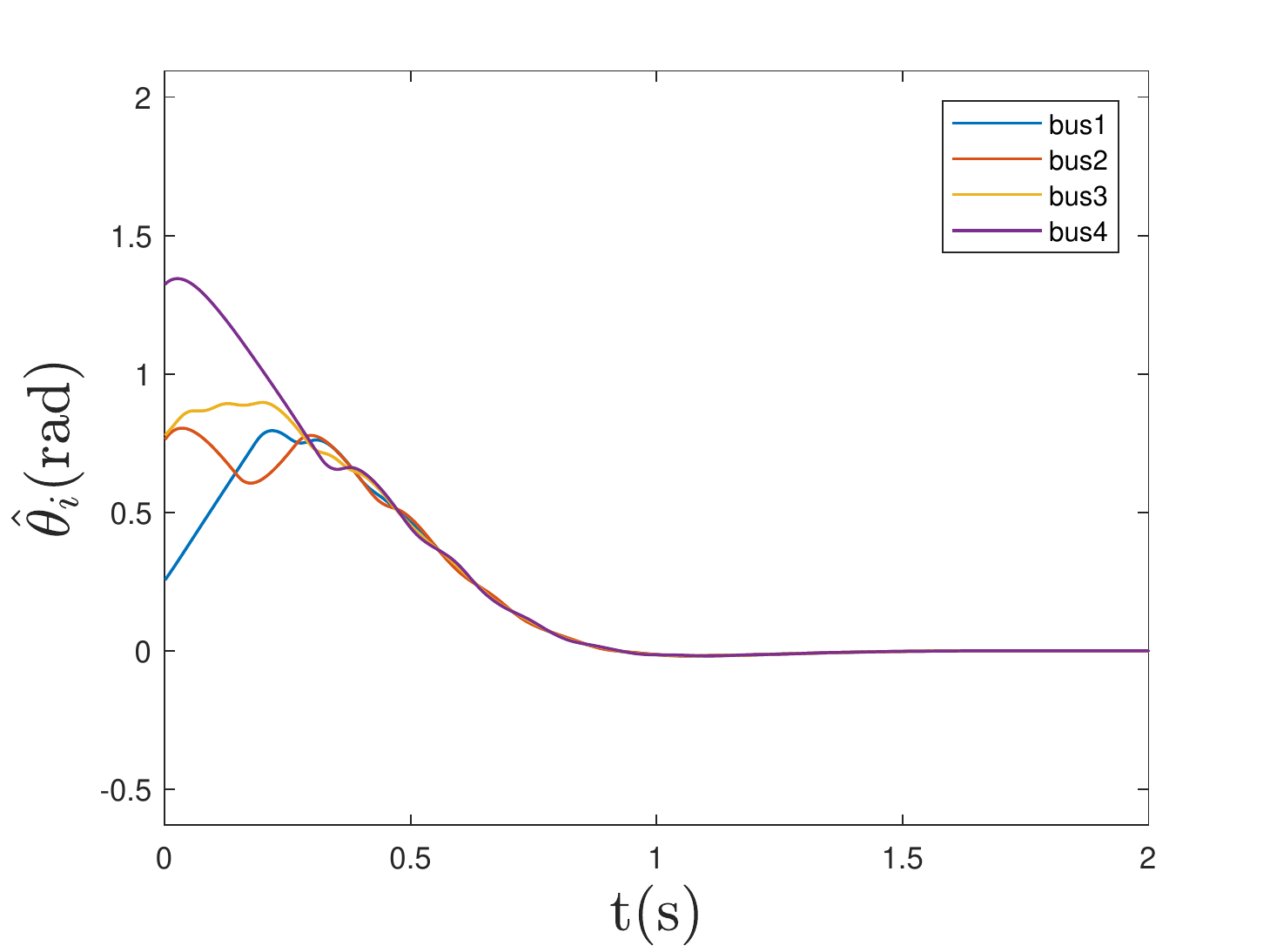}
		\hspace{100in}\vspace{-3mm}
		\caption{$\hat{\theta}_i$ of bus1, bus2 ,bus3 and bus4.}
	\end{center}
\end{figure}
\begin{figure}[!htb]
	\begin{center}
		\includegraphics[width=3in]{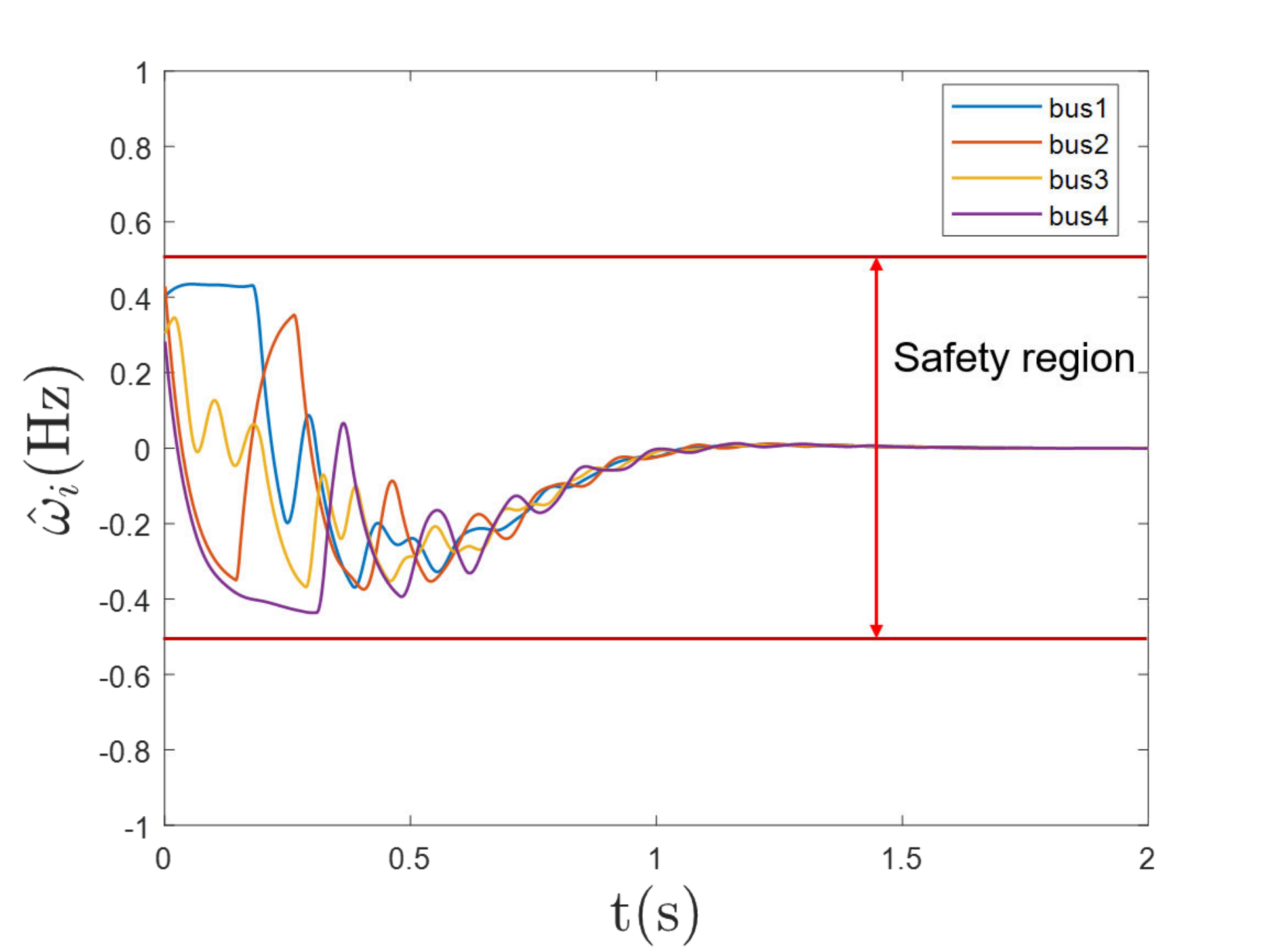}
		\hspace{100in}\vspace{-3mm}
		\caption{$\hat{\omega}_i$ of bus1, bus2 ,bus3 and bus4.}
	\end{center}
\end{figure}

\section{Conclusions}
In this paper, we have developed a novel control approach for the sparse and safe frequency regulation in MGs with inverters. The framework of this control design is comprised of three steps. First, the SLFC is designed and acts as a nominal control. Second, a family of conditions for safe control are designed using CBFs. Finally, by unifying SLFC and CBFs, a QP problem is constructed, and the real-time control is obtained by solving the QP problem.  Importantly, we have also found that the real-time control for each inverter depends on the cross-layer communication network topology which is the union of the one between controllers from SLFC and the one determined by the power flow network. In addition, our approach has been verified by the case study.

%\section*{Acknowledgment}

%The preferred spelling of the word ``acknowledgment'' in %America is without 
%an ``e'' after the ``g''. Avoid the stilted expression ``one %of us (R. B. 
%G.) thanks $\ldots$''. Instead, try ``R. B. G. %thanks$\ldots$''. Put sponsor 
%acknowledgments in the unnumbered footnote on the first page.

\bibliographystyle{IEEEtran}
\bibliography{mybibfile}
\end{document}